\documentclass[aps,pra,10pt,groupedaddress,twocolumn,showpacs]{revtex4-1}

\usepackage[english]{babel}
\usepackage[dvips]{graphicx}
\usepackage{indentfirst}
\usepackage{tabularx}
\usepackage[normalem]{ulem}
\usepackage{amsmath}
\usepackage{ulem}

\begin{document}
\title{Variational Monte Carlo study of soliton excitations in hard-sphere Bose gases}
\author{R. Rota}
\affiliation{Laboratoire Mat\'eriaux et Ph\'enom\`enes Quantiques, Universit\'e Paris Diderot-Paris 7, 75205 Paris Cedex 13, France}
\author{S. Giorgini}
\affiliation{Dipartimento di Fisica, Universit\`a di Trento and INO-CNR BEC Center, 38123 Povo, Trento, Italy}
\date{\today}

\begin{abstract}
By using a full many-body approach, we calculate the excitation energy, the effective mass and the density profile of soliton states in a three dimensional Bose gas of hard spheres at zero temperature. The many-body wave function used to describe the soliton contains a one-body term, derived from the solution of the Gross-Pitaevskii equation, and a two-body Jastrow term which accounts for the repulsive correlations between atoms. We optimize the parameters in the many-body wave function via a Variational Monte Carlo procedure, calculating the grand-canonical energy and the canonical momentum of the system in a moving reference frame where the soliton is stationary. As the density of the gas is increased, significant deviations from the mean-field predictions are found for the excitation energy and the density profile of both dark and grey solitons. In particular, the soliton effective mass $m^\ast$ and the mass $m\Delta N$ of missing particles in the region of the density depression are smaller than the result from the Gross-Pitaevskii equation, their ratio being however well reproduced by this theory up to large values of the gas parameter. We also calculate the profile of the condensate density around the soliton notch finding good agreement with the prediction of the local density approximation.   
\end{abstract}


\maketitle

\section{Introduction}

Solitons are nonlinear collective excitations which appear in a wide range of physical systems, from classical fluids to optical fibers: they are characterized as localized wave forms which travel in a uniform medium at a constant velocity without spreading. Ultracold atomic gases are well-controlled quantum systems which are particularly suitable for the investigation of solitons. Dark and grey solitons (i.e. localized density depressions in a homogeneous background) can indeed be produced in repulsive Bose-Einstein condensates (BECs) by imprinting a phase jump on the atomic cloud~\cite{Burger99,Denschlag00,Anderson01,Becker08}, by density engineering~\cite{Dutton01,Shomroni09} or by merging two coherent BECs initially prepared in a double well potential~\cite{Weller08}. Furthermore, other kinds of solitons (e.g. bright solitons, gap solitons and dark-bright solitons) have also been created and detected in bosonic quantum gases~\cite{Strecker02,Khaykovich02,Eiermann04,Becker08}.

From the theoretical point of view, soliton excitations in superfluids are studied mostly within mean-field approaches based on the description of the system in terms of a complex order parameter which evolves in space and time according to a nonlinear equation. In the case of Bose superfluids the paradigmatic theory is provided by the Gross-Pitaevskii (GP) equation where soliton solutions have been widely investigated~\cite{BECBook}. A conceptually similar, even though technically more involved, mean-field approach exists also for Fermi superfluids and is based on the time-dependent Bogoliubov-de Gennes equations~\cite{Antezza07,Liao11,Scott11,Spuntarelli11}. We point out that mean-field solitons are stable excitations in one dimension (1D), where the phase of the order parameter changes sign at a single point. In higher dimensions they can be stabilized by a sufficiently strong confinement which reduces the soliton nodal surface, but otherwise undergo a snake instability towards the formation of vortices or vortex rings~\cite{Muryshev99,Anderson01,Cetoli13}. 

An important question is to understand how the properties of soliton excitations change when interparticle interactions and/or correlations increase and the mean-field picture in terms of the order parameter eventually fails. Using time-dependent Bogoliubov theory~\cite{Dziarmaga02A1,Dziarmaga02A2,Law03} as well as more sophisticated numerical techniques~\cite{Mishmash09Lett,Mishmash09A,Martin10,Rubbo12,Delande14} to treat the many-body dynamics of 1D bosons at zero temperature, it has been shown that beyond mean-field effects tend to deplete the condensate and to fill the soliton notch making dark solitons unstable. To our best knowledge, however, no theoretical studies have addressed the role of many-body correlations in determining relevant properties of grey solitons in a three-dimensional (3D) Bose gas, such as their excitation energy, density profile and effective mass. In particular, the problem of the value of the effective mass beyond the limits of mean-field theory became prominent in the first interpretation of the recent experiment~\cite{Zwierlein13}, where localized excitations obtained by phase imprinting in an ultracold Fermi gas of $^6$Li were observed to move much more slowly than predicted by mean-field theory. These defects were initially ascribed as solitons, but lately a more precise imaging technique revealed that the moving defects were solitonic vortices, thus explaining the puzzle of their greater inertia~\cite{Zwierlein14}. Nevertheless, the dependence of the soliton effective mass on the strength of interactions as the gas gets less dilute remains an open question worth considering.

In this paper we perform full many-body simulations of soliton excitations in a 3D Bose gas at zero temperature. We calculate the energetic and structural properties of grey and dark solitons, both in the very dilute and in the strongly interacting regime, and we determine their effective mass from the excitation energy vs. speed dependence. The results are systematically compared with the predictions of GP equation and important deviations are found as the density is increased. Remarkably, the ratio $\frac{m^\ast}{m\Delta N}$ between the soliton effective mass and the mass of missing particles in the density dip, which is the crucial parameter in the dynamics of solitons as macroscopic localized defects, is found to remain close to the mean-field value up to very high densities.

The paper is organized as follows. In Sec. \ref{sec:method}, we introduce the variational approach used in the numerical determination of the properties of the soliton: we discuss the form of the many-body wave function and the energy functional in terms of which the soliton state can be defined. In Sec. \ref{sec:results}, we present the results obtained within this many-body approach and we discuss the appearance of corrections beyond the mean-field description of solitons. Finally, in Sec. \ref{sec:conclusions}, we draw our conclusions.

\section{Numerical Method}\label{sec:method}

The system of $N$ interacting identical bosons of mass $m$ is described by the Hamiltonian of the quantum degenerate hard-sphere (HS) model, that is 
\begin{equation}
H=-\frac{\hbar^2}{2m}\sum_{i=1}^N \nabla_i^2 + \sum_{i<j}V(r_{ij}) \;,
\label{Hamiltonian}
\end{equation}
where $r_{ij}$ is the distance between the $i$-th and $j$-th particle and the central potential $V$ is modeled by the hard-sphere interaction, 
\begin{eqnarray}\label{eq:HSpotential}
V(r) = \left\{
\begin{array}{cll}
\infty & (r \le a) \\
0 & (r > a) \;,
\end{array}
\right.
\end{eqnarray}
in terms of the $s$-wave scattering length $a$. This model has been widely used to describe quantum many-body systems with short-range repulsive interaction, both in the weakly and in the strongly interacting regime. Indeed, it is able  not only to capture the essential properties of dilute systems like ultracold gases with positive scattering-length, for which the details of the interatomic potential are irrelevant, but also to characterize semi-quantitatively the static properties of strongly interacting systems like superfluid $^4$He \cite{Huang57,Hansen71,Kalos74}. Moreover, the choice of the HS model is particularly convenient as it allows to parametrize the strength of the interactions by just varying the value of the dimensionless gas parameter $na^3$, where $n=N/V$ is the density of the gas.

To describe soliton excitations, we consider states of the gas described by the many-body wave function
\begin{equation}
\Psi_S({\bf r}_1,\ldots,{\bf r}_N;t) = {\cal{A}}_S\prod_{i=1}^N \phi(z_i - v t) \prod_{i<j} f(r_{ij}) \;,
\label{ansatz}
\end{equation}
where ${\cal{A}}_S$ is a normalization factor. The product of two-body Jastrow terms accounts for interparticle correlations: we model them using the positive function $f(r)=\frac{\sin[k(r-a)]}{r}$ in the interval $a<r<R_M$, that is the exact solution of the scattering problem for a hard-sphere potential. For $r > R_M$, we match $f(r)$ with the constant $\frac{\sin[k(R_M-a)]}{R_M}$. This function satisfies the boundary condition $f(r)=0$ at $r<a$ and the wave vector $k$ is determined from the condition $f^\prime(r=R_M)=0$ of the first derivative $f^\prime$ at the matching point $R_M$. The value of $R_M$ is used as a variational parameter and is optimized to minimize the excitation energy of the dark soliton. The complex one-body term $\phi$ is given by 
\begin{equation}
\phi(z) = \sqrt{1-\alpha^2} \tanh \left(\frac{z}{\gamma \xi} \sqrt{1-\alpha^2} \right) +i\alpha \;,
\label{onebodyterm}
\end{equation}
and describes a perturbation in the density profile propagating along the $z$-axis with velocity $v$. 
Here $\xi = 1/\sqrt{8 \pi n a}$ is the healing length of the gas at the background density $n$, while $\alpha$ and $\gamma$ are two variational parameters. The functional form of $\phi$ is dictated from the solution $\sqrt{n}\phi(z-vt)$ of the GP equation describing dark and grey solitons \cite{BECBook}: in this case, the values of the parameters are $\gamma = \sqrt{2}$ and $\alpha = v/c_0$, with $c_0=\frac{\hbar}{\sqrt{2}m\xi}$ the speed of sound within the GP theory. The phase of $\phi$ undergoes the finite change $\Delta S =2 \arccos(\alpha)$  as $z$ varies from $+\infty$ to $-\infty$, whereas the density perturbation is localized around the origin of the $z$-axis on a length scale fixed by $\gamma\xi/\sqrt{1-\alpha^2}$. 
The wave function (\ref{ansatz}) is translationally invariant in the $xy$ plane and its global phase is the sum of single-particle contributions each of which changes sign at the $z_i-vt=0$ plane. 

Since we assumed that the time $t$ enters the many-body wave function only through the linear combination $z_i - v t$, it is convenient to describe the problem in a moving reference frame, where the soliton is stationary. In analogy with the GP equation we define soliton states as stationary points of the functional
\begin{equation}
\Omega[\Psi_S] = E[\Psi_S] - v P_C[\Psi_S] \ ,
\label{functional}
\end{equation}
where $E$ is the grand-canonical energy of the many-body system, 
\begin{equation}
E[\Psi_S] = \int d\tilde{\bf R}\; \Psi_S^\ast(\tilde{\bf R})(H-\mu)\Psi_S(\tilde{\bf R}) \ ,
\label{eq:grand-canonical-energy}
\end{equation}
and $P_C$ is the canonical momentum 
\begin{eqnarray}
P_C[\Psi_S]= \hbar n L_xL_y(\Delta S -\pi) &-& \frac{i\hbar}{2}\int d\tilde{\bf R}
\label{pcanonical}\\
\times \left[ \Psi_S^\ast(\tilde{\bf R})\sum_{i=1}^N\frac{d}{d\tilde{z}_i}\Psi_S(\tilde{\bf R}) \right.
&-& \left. \Psi_S(\tilde{\bf R})\sum_{i=1}^N\frac{d}{d\tilde{z}_i}\Psi_S^\ast(\tilde{\bf R})\right] \;.
\nonumber
\end{eqnarray}
The set of particle coordinates $\tilde{\bf R}=(\tilde{\bf r}_1,\dots,\tilde{\bf r}_N)$ refers to the moving reference frame $\tilde{\bf r}_i=(x_i,y_i,\tilde{z}_i)$, where $\tilde{z}_i=z_i-vt$. Furthermore, $L_x$ and $L_y$ indicate the length of the system respectively in the $x$ and $y$ directions. In Eq.~(\ref{eq:grand-canonical-energy}), $\mu$ is the chemical potential of the homogeneous gas enforcing the boundary condition that the soliton state reaches the asymptotic density $n$ when $\tilde{z}_i=\pm\infty$ for all the particles. The first term in Eq.~(\ref{pcanonical}) arises instead from the boundary condition that the many-body wave function $\Psi_S$ has a phase jump $N\Delta S$ across the soliton and it is necessary to account for the motion of the background density in the moving reference frame~\cite{Shevchenko88,BECBook,Scott11}. The functional $\Omega$ can be derived from the principle of minimal action applied to the quantity 
\begin{equation}
A=\int dt\int d{\bf R}\Psi_S^\ast({\bf R},t)(-i\hbar\partial_t+H-\mu)\Psi_S({\bf R},t)
\end{equation}
where $\Psi_S$ is a state of the form (\ref{ansatz}) complying with the boundary conditions explained above~\cite{note}. 

The simulations for a given value of the gas parameter $na^3$ are carried out using the Variational Monte Carlo (VMC) technique. As a first step, we calculate the properties of the uniform ground state by performing VMC simulations of $N_U$ particles in a box of volume $V = N_U/n$ with periodic boundary conditions. We use the function $\Psi_U({\bf R}) = {\cal{A}}_U\prod_{i<j} f(r_{ij})$ to describe the normalized ground state and we determine the corresponding energy $E_0$ and chemical potential $\mu=\left(\frac{dE_0}{dN}\right)_V$. The soliton state is simulated in the same box of volume $V=L_xL_yL_z$, using a number $N_S < N_U$ of particles chosen so as to comply with the boundary condition of the density profile reaching the uniform value $n$ far away from the soliton plane. The volume of the simulation box is chosen large enough to prevent finite-size effects: typical values are $L_z\gtrsim20\xi$ and $L_x=L_y\gtrsim8\xi$. Moreover, the value $R_M$ of the matching point in the Jastrow factor is kept the same for the calculation of the $\Psi_U$ and $\Psi_S$ state. For each value of the velocity $v$ of the soliton (given as an input parameter), we calculate $\Omega$ as a function of the variational parameters and we look for the stationary points to determine the optimal values of $\alpha$ and $\gamma$. In analogy with the GP equation, the soliton state corresponds to a saddle point of $\Omega$, {\it i.e.} a minimum as a function of $\gamma$ and a maximum as a function of $\alpha$.

\begin{figure}
\includegraphics[angle=-90,width=0.44\textwidth]{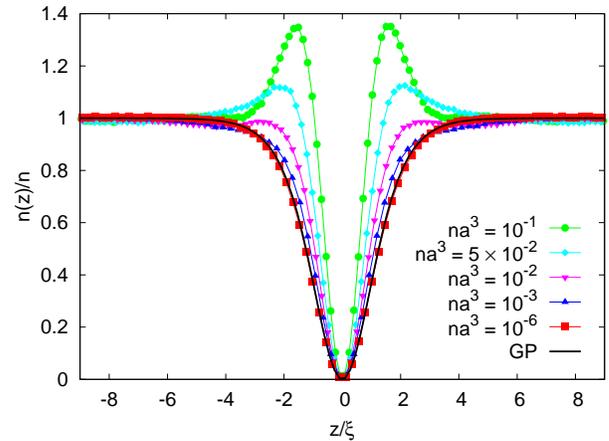}
\caption{(Color online) Density profile of a dark soliton for different values of the gas parameter (lines are a guide to the eyes). The solution of the GP equation is also shown for comparison.}
\label{fig:profile_na3}
\end{figure}

\begin{figure}
\includegraphics[angle=-90,width=0.44\textwidth]{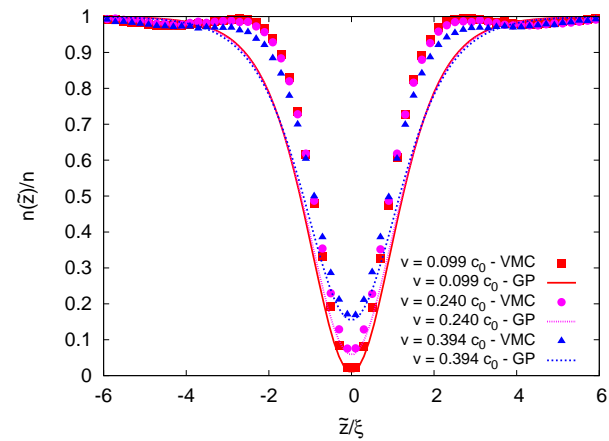}
\caption{(Color online) Density profile of the soliton at  the density $na^3 = 10^{-2}$ for different velocities in the reference frame where the soliton is at rest. The mean-field curves are also shown.}
\label{fig:profile_v}
\end{figure}

\section{Results}\label{sec:results}

We start the discussion of the structural properties by considering the limit of vanishing velocity, $v/c_0\to0$, where the optimal value of $\alpha$ approaches zero and the wave function $\Psi_S$ describes a dark soliton with a phase jump $\Delta S=\pi$. The density profiles $n(z)$ of these stationary excitations are shown in Fig.~\ref{fig:profile_na3} for different values of the gas parameter and compared with the profile obtained within the GP approach. A systematic analysis of the effects of many-body interactions on the condensate wave function in time-dependent problems has been performed in Ref. \cite{Castin98}. In this work, it has been shown that the relevant parameter for the deviations from the solution of GP equation is the square root of the quantum depletion, which scales as $(na^3)^{1/4}$. In Fig.~\ref{fig:profile_na3}, we can see that the VMC results are in excellent agreement with mean-field predictions at the lowest density, $na^3 = 10^{-6}$. As the gas parameter increases, we notice some deviations from the solution of GP equation although they become significative only at $na^3>10^{-3}$. At these higher densities the width of the soliton, in units of the healing length, decreases and some oscillations appear in the density profile. These oscillations are more pronounced at the largest densities and get damped in regions far from the soliton. Similar oscillations appear in the density profile of liquid $^4$He around a vortex\cite{Dalfovo92} and their appearance can be interpreted as the tendency of the system to organize itself in shells of atoms around the defect. We notice that, for all values of the gas parameter, the density vanishes on the $z=0$ soliton plane in agreement with the GP prediction for dark solitons. This is a consequence of the choice (\ref{onebodyterm}) of the one-body term which, as $\alpha\to0$, generates for each particle a phase discontinuity at the $z=0$ plane. 

In Fig.~\ref{fig:profile_v}, we show the density profile of grey solitons moving with different velocities $v$ in a gas of background density $na^3=10^{-2}$, which corresponds to a regime of relatively strong interactions among the particles. One observes clear deviations from the GP profile in the region of the wings of the soliton, similar to the $v = 0$ case (see Fig.~\ref{fig:profile_na3}). The discrepancies between the two profiles are less pronounced in the central region of the soliton where the VMC results at all velocities are only slightly above the GP predictions. 

\begin{figure}
\includegraphics[angle=-90,width=0.40\textwidth]{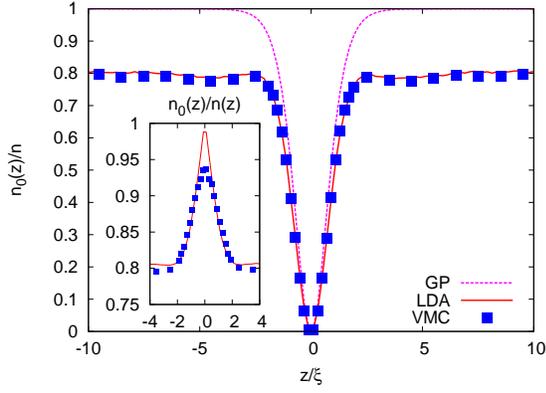}
\caption{(Color online) Condensate density profile $n_0(z)$ of a dark soliton at $na^3 = 10^{-2}$. The results of both GP equation and LDA are also shown. Inset: local condensate fraction $n_0(z)/n(z)$ compared with the LDA result.}
\label{fig:condensate}
\end{figure}

Another important quantity accessible in our microscopic approach is the condensate profile $n_0(z)$ around the defect. This can be obtained from the long-distance behavior of the one-body density matrix in the $xy$ plane at a fixed value of the $z$ coordinate:
\begin{equation}
n_0(z)=\lim_{|{\bf r}_1^\prime-{\bf r}_1|\to\infty}  \int d{\bf R}_{N-1} \, |\Psi_S|^2 \frac{\Psi_S({\bf r}_1^\prime,{\bf r}_2,\dots,{\bf r}_N)}{\Psi_S({\bf r}_1,{\bf r}_2,\dots,{\bf r}_N)} \;,
\label{n0}
\end{equation}
with ${\bf r_1^\prime}=(x_1^\prime,y_1^\prime,z)$, ${\bf r_1}=(x_1,y_1,z)$ and ${\bf R}_{N-1} = \{{\bf r}_2, \ldots, {\bf r}_N\}$. In Fig.~\ref{fig:condensate} we show the condensate profile of a dark soliton ($v=0$) at the density $na^3=10^{-2}$. Far away from the soliton plane at $z=0$, the condensate density reaches the bulk value $n_0\simeq0.8n$ in agreement with the result for a homogeneous gas~\cite{Barca99}. When approaching the soliton plane, we find that $n_0(z)$ is in closer agreement with the GP profile than the total density $n(z)$.  Remarkably, the condensate profile is well reproduced by the local density approximation (LDA) where we determine $n_0(z)$ from the results of the quantum depletion in a homogeneous gas~\cite{Barca00,Rossi13} at the local density $n(z)$.  Only in the region very close to the soliton plane, the LDA result shows deviations on the order of 5\% with respect to the local condensate fraction $n_0(z)/n(z)$ obtained from VMC calculations (see inset of Fig.~\ref{fig:condensate}).

\begin{figure}
\includegraphics[angle=-90,width=0.43\textwidth]{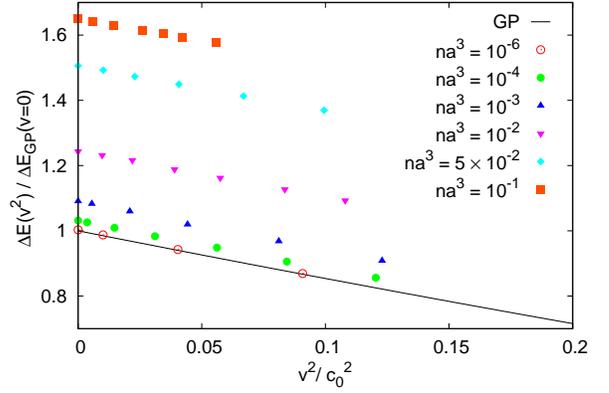}
\caption{(Color online) Excitation energy $\Delta E$ of the soliton as a function of $(v/c_0)^2$. The solid line corresponds to the GP result. Statistical errors are below symbol size.}
\label{fig:ener_v2}
\end{figure}

\begin{figure}
\includegraphics[angle=-90,width=0.40\textwidth]{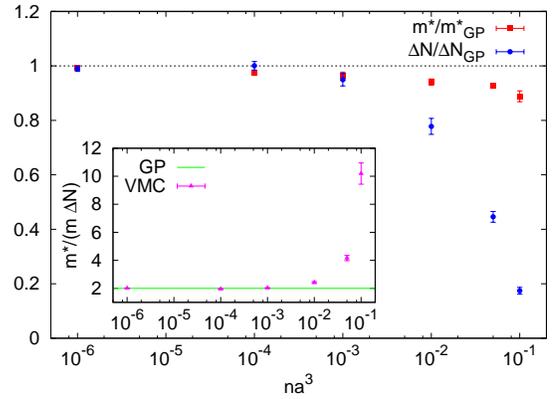}
\caption{(Color online) Ratio of the VMC to the GP effective mass $m^\ast$ and mass of missing particles in the soliton dip $m\Delta N$ as a function of the gas parameter.  Inset: ratio $\frac{m^\ast}{m\Delta N}$ as a function of $na^3$.}
\label{fig:effectivemass}
\end{figure}

The energetics of solitons is reported in Fig.~\ref{fig:ener_v2}, where we show the excitation energy $\Delta E$ of the soliton as a function of $v^2$ and for different values of the gas parameter. This value is obtained from the difference $\Delta E = E[\Psi_S] - E[\Psi_U]$ of the grand-canonical energy (\ref{eq:grand-canonical-energy}) between the states with and without the soliton. At the smallest value of the gas parameter ($na^3=10^{-6}$), we find a good agreement between the VMC result and the GP prediction 
\begin{equation}
\Delta E_{GP}= L_x L_y \frac{4}{3} \hbar c_0 n\left(1- \frac{v^2}{c_0^2} \right)^{3/2} \ . 
\end{equation}
Significantly larger excitation energies are obtained at higher densities, showing that beyond mean-field effects result in a more pronounced enhancement of the soliton energy $E[\Psi_S]$ compared to the increase of the ground-state energy $E[\Psi_U]$ of the background homogeneous gas~\cite{Barca99}. From the slope of $\Delta E$ with respect to $v^2$ we extract the effective mass $m^\ast$ of the soliton shown in Fig.~\ref{fig:effectivemass} as a function of $na^3$. Despite the fact that the excitation energy becomes larger as the density increases, the effective mass per surface unit remains always close to the mean-field prediction $m_{GP}^\ast=-4\frac{\hbar n}{c_0}$ with a reduction of only $\sim$10\% at the largest density. We would like to point out that values of $m^\ast$ consistent with the ones reported in Fig.~\ref{fig:effectivemass} are obtained from the linear dependence of the canonical momentum $P_C$ in Eq.~\ref{pcanonical} on the velocity $v$ of the soliton.

In Fig.~\ref{fig:effectivemass} we also show the mass $m\Delta N$ of missing particles in the notch of a dark soliton per unit surface, obtained from the formula $\Delta N =(N_S - N_U)/L_x L_y$. We find that up to densities $na^3\simeq10^{-3}$, the suppression of $\Delta N$ with respect to the GP value $\Delta N_{GP} = -2\frac{\hbar n}{mc_0}$ is small and approximately equal to the reduction of the effective mass. At higher densities, the ratio $\Delta N/\Delta N_{GP}$ drops down, following the filling of the soliton dip as shown in Fig.~\ref{fig:profile_na3}. Due to the qualitatively similar behavior of $m^\ast$ and $m\Delta N$, their ratio remains surprisingly close to the mean-field prediction $\frac{m_{GP}^\ast}{m\Delta N_{GP}}=2$ up to large values of the gas parameter ($na^3\simeq10^{-2}$). Only at the largest density $na^3 = 10^{-1}$ this ratio shoots up because of the large suppression in $\Delta N$ (see inset of Fig.~\ref{fig:effectivemass}). It is worth stressing that the ratio $m^*/(m\Delta N)$ is an important parameter characterizing the dynamics of solitons as localized objects, being in particular related to the frequency of their oscillation in a harmonic confinement~\cite{Scott11}.

\section{Conclusions}\label{sec:conclusions}

We have investigated the structural and energetics properties of soliton excitations in 3D using for the first time a fully microscopic, many-body approach and carefully comparing our results  with standard GP theory. We provide quantitative predictions for the effective mass and the mass of missing particles as a function of the gas parameter, showing that their ratio (that is the key parameter in the dynamics of the solitons in a harmonic confinement) is in good agreement with mean-field prediction even for regimes where the time-dependent Gross-Pitaevskii equation usually fails.

In particular, the results at high density can be relevant for future experiments on the dynamics of solitons in strongly interacting composite bosons realized on the BEC side of a Feshbach resonance in a two component Fermi gas. At the moment, the main hindrance to the experimental determination of $m^*/(m \, \Delta N)$ is the fast decay of the solitons, which limits the possibility of measuring the period of oscillation of these defects in harmonically trapped gases ~\cite{Zwierlein15}. By applying a stronger radial confinement to cigar-shaped configurations, it might be possible to increase the lifetime of solitons and thus to study their dynamics in regimes where the gas parameter is large.

\section*{Acknowledgments}

Useful discussions with L. P. Pitaevskii, F. Dalfovo and G. Astrakharchik are gratefully acknowledged. This work has been supported by ERC through the QGBE grant.


\begin{thebibliography}{50}
\bibitem{note} One should note that the relation between the wave functions in the two reference frames is given by $\Psi_S({\bf R},t)=\Psi_S(\tilde{{\bf R}})\exp(-i\frac{Nmv^2}{2\hbar}t)\exp(-i\frac{mv}{\hbar}\sum_i\tilde{z}_i)$.
\end{thebibliography}
\end{document}